\documentstyle[12pt,a4wide,epsfig,amssymb]{article}

\def\MBp{M_{\tilde{B}'}}
\def\MWp{M_{\tilde{W}'_3}}
\def\Bp{\tilde{B}'}
\def\Wp{\tilde{W}'_3}
\title{
\vspace*{-1.0cm}
\begin{flushright}
{\normalsize LNF-00/017(P)\\[-2pt]
UCRHEP-T278\\[-14pt]
}
\end{flushright}
\vspace{1.3cm}
{\Large Leptogenesis from Neutralino Decay
with Nonholomorphic R-Parity Violation} 
\vspace*{0.8cm}
}
\author{Thomas Hambye$^{a\,}$, Ernest Ma$^{b\,}$, 
and Utpal Sarkar$^{b,c\,}$\\[0.5cm]
\normalsize $a$: {\it INFN - Laboratori Nazionali di Frascati,
P.O. Box 13, I-00044 Frascati, Italy}\\[1mm] 
\normalsize $b$: {\it Department of Physics, University of California,
          Riverside, California 92521, USA}\\
\normalsize $c$: {\it Physical Research Laboratory, 
Ahmedabad 380 009, India}\\[2cm] 
}
\date{}
\begin{document}
\maketitle
\thispagestyle{empty}
\begin{abstract}

In supersymmetric models with lepton-number violation, hence also R-parity 
violation, it is easy to have realistic neutrino masses, but then leptogenesis 
becomes difficult to achieve.  After explaining the general problems involved, 
we study the details of a model which escapes these constraints and generates 
a lepton asymmetry, which gets converted into the present observed baryon 
asymmetry of the Universe through the electroweak sphalerons.  This model 
requires the presence of certain nonholomorphic R-parity violating terms.  
For completeness we also present the most general R-parity violating 
Lagrangian with soft nonholomorphic terms and study their consequences for 
the charged-scalar mass matrix. New contributions to neutrino masses in this 
scenario are discussed.

\end{abstract}
\vspace*{\fill}
\noindent
PACS numbers: {12.60.Jv, 11.30.Fs, 98.80.Cq, 14.80.Ly}
\newpage
%
\section{Introduction}

The creation of a lepton asymmetry, i.e. leptogenesis \cite{fy,lepto,ht}, 
which gets converted into the present observed baryon asymmetry of the 
Universe, is closely related to the mechanism by which neutrinos obtain 
mass.  In general, all models of Majorana neutrino masses with the same 
low-energy particle content as that of the standard model are equivalent 
in the sense that they are all characterized by the same nonrenormalizable 
dimension-five operator \cite{wein} $\Lambda^{-1} \nu_i \nu_j \phi^0 \phi^0$. 
Different models of neutrino mass are merely different realizations \cite{ma} 
of this operator.  They become distinguishable only at high energies, and 
since their interactions must violate lepton number, leptogenesis is a very 
natural possibility.  For the canonical seesaw mechanism \cite{seesaw} and 
the Higgs triplet model \cite{triplet}, leptogenesis does indeed occur 
naturally \cite{fy,triplet}.  On the other hand, if neutrino masses are 
obtained radiatively \cite{ma}, not only is leptogenesis difficult to 
achieve, the mechanism by which the former is accomplished leads naturally 
to the erasure of any primordial baryon asymmetry of the Universe 
\cite{fy1,erase, rad}.  This is especially true in supersymmetric models of 
neutrino mass \cite{numass} with R-parity violation.  In a recent article 
\cite{hms}, it was pointed out that leptogenesis is still possible in this 
case, provided that certain conditions regarding the R-parity violating 
terms are satisfied.  Here we study this model in detail.

In Section 2 we write down the superpotential of the lepton-number 
violating (but baryon-number conserving) extension of the supersymmetric 
standard model, together with all possible soft supersymmetric breaking 
terms, including the nonholomorphic terms \cite{hr}.  In Section 3 we 
consider bilinear R-parity violation and how leptogenesis is related to 
neutrino mass in this limited scenario.  We find it to be negligible for 
realistic values of $m_\nu$.  In Section 4 we discuss how leptogenesis may 
occur without being constrained by neutrino mass in an expanded scenario.  
We assume negligible (enhanced) mixing between doublet (singlet) sleptons and 
charged Higgs bosons by allowing nonholomorphic soft supersymmetry breaking 
terms.  In Section 5 we present the details of our calculations using the 
Boltzmann equations for obtaining the eventual lepton asymmetry.  In Section 
6, the complete charged-scalar mass matrix is displayed and analyzed.  In 
Section 7, a new two-loop mechanism for neutrino mass is proposed.  Finally 
in Section 8, there are some concluding remarks.

%
%
\section{Superpotential and Soft Supersymmetry Breaking \label{softterms}}
In an unrestricted supersymmetric extension of the standard model of particle 
interactions, the chiral scalar superfields allow baryon-number violating 
terms which are not necessarily suppressed.  These dangerous terms are 
usually avoided by assuming a conserved discrete quantum number for each 
particle called R-parity, which is defined as
\begin{equation}
{\rm R} \equiv (-1)^{3B+L+2J},
\end{equation}
where $B$ is its baryon number, $L$ its lepton number, and $J$ its spin 
angular momentum.  With this definition, the standard-model particles 
have R = +1 and their supersymmetric partners have R = $-1$.  We can
list the three families of leptons and quarks of the standard model
using the notation where all superfields are considered left-handed:
\begin{eqnarray}
&& L_i = (\nu_i, e_{i_{\tiny L}}) 
\sim (1,2,-1/2), ~~~ e^c_i \sim (1,1,1), \\
&& Q_i = (u_i, d_i) \sim (3,2,1/6), \\
&& u^c_i \sim (3^*,1,-2/3), ~~~ d^c_i \sim (3^*,1,1/3),
\end{eqnarray}
where $i$ is the family index, and the two Higgs doublets are given by
\begin{eqnarray}
&& H_1 = (h_1^0,h_1^-) \sim (1,2,-1/2), \\
&& H_2 = (h_2^+,h_2^0) \sim (1,2,1/2),
\end{eqnarray}
where the $SU(3)_C \times SU(2)_L \times U(1)_Y$ content of each superfield 
is also indicated.  If R-parity is conserved, the superpotential is 
restricted to have only the terms
\begin{equation}
W = \mu H_1 H_2 + f_{ij}^e H_1 L_i e^c_j + f^d_{ij} H_1 Q_i d^c_j 
+ f^u_{ij} H_2 Q_i u^c_j.
\label{superpot}
\end{equation}
In this case, both baryon and lepton numbers are conserved.  However, to 
forbid proton decay, it is sufficient to conserve either baryon number or 
lepton number (because the final state of the proton decay must contain 
a lepton or antilepton).  If only baryon number or only lepton number is 
violated (thus R-parity is also violated), the conservation of the other 
quantum number is enough to satisfy all present experimental constraints. 
This has motivated numerous studies of R-parity violating models. 

If R-parity is violated, the superpotential contains the additional terms
\begin{equation}
W' = \mu_i L_i H_2 + \lambda_{ijk} L_i L_j e^c_k + \lambda'_{ijk} 
L_i Q_j d^c_k + \lambda''_{ijk} u_i^c d_j^c d_k^c .
\end{equation}
We cannot have all of these terms because then the proton will decay very 
quickly.  We may choose only the lepton-number violating terms or only the 
baryon-number violating terms.  Following the overwhelming choice of many 
others, we consider only the former case and set $ \lambda''_{ijk} = 0$. 
The remaining terms may now induce nonzero neutrino masses, either from 
mixing with the neutralino mass matrix, or in one-loop order\cite{numass}. 
Although these terms are allowed, we do not know if they originate from 
any fundamental theory, so the couplings are considered free parameters, 
constrained only by experiment.

Other free parameters exist in the minimal supersymmetric standard 
model (MSSM), i.e. the soft supersymmetry breaking terms, which do not 
introduce quadratic divergences to the renormalized theory.  Usually only the 
holomorphic terms are considered which come from the chiral superpotential 
interacting with gravity, together with the gaugino masses. The most general 
such Lagrangian conserving R-parity is:
\begin{eqnarray}
{\cal L}_{soft}&=&
- \tilde{L}_i^{a \ast} (M_L^2)_{ij} \tilde{L}_j^a 
- \tilde{e}_i^{c \ast} (M_e^2)_{ij} \tilde{e}_j^c
- \tilde{Q}_i^{a \ast} (M_Q^2)_{ij} \tilde{Q}_j^a  
- \tilde{u}_i^{c \ast} (M_u^2)_{ij} \tilde{u}_j^c 
\nonumber\\[1.5mm] 
&&
- \tilde{d}_i^{c \ast} (M_d^2)_{ij} \tilde{d}_j^c 
- M_{H_1}^2 H_1^{a \ast} H_1^a - M_{H_2}^2 H_2^{a \ast} H_2^a 
- \varepsilon_{ab} ( B H_1^a H_2^b  +  h.c.) 
\nonumber\\[1mm] 
&&
-\varepsilon_{ab} \Big( (A_e f_e)_{ij} H_1^a \tilde{L}_i^b \tilde{e}_j^c 
+ (A_u f_u)_{ij} H_2^b \tilde{Q}_i^a \tilde{u}_j^c 
+(A_d f_d)_{ij} H_1^a \tilde{Q}_i^{b} \tilde{d}_j^c \, + \, h.c. 
\Big)
\nonumber\\[1mm] 
&&
-\frac{1}{2} \Big( M_3 \tilde{g} \tilde{g} 
+ M_2 \tilde{W} \tilde{W}
+ M_1 \tilde{B} \tilde{B} \, + \, h.c. \Big)
\label{lsoft}\,.
\end{eqnarray}
If R-parity is violated, more soft terms may be present, i.e.
\begin{equation}
{\cal L}_{soft}^{R \! \! \! /}=  - \varepsilon_{ab} \big( 
{B}_i' \tilde{L}_i^a H_2^b
+ {A}^{\prime e}_{ijk} \tilde{L}_i^a \tilde{L}_j^b \tilde{e}_k^c
+ {A}^{\prime d}_{ijk} \tilde{Q}_i^a \tilde{L}_j^b \tilde{d}_k^c \big)
- {A}^{\prime {\tiny S}}_{ijk} \tilde{u}_i^c \tilde{d}_j^c \tilde{d}_k^c
\, + \, h.c. 
\label{lsoftRviol}\,.
\end{equation}
We follow the convention that the coupling constants of all the 
R-parity conserving soft terms are denoted without a prime, while the R-parity
violating terms are denoted with a prime.

Since the soft terms may originate from gravity couplings, there is no clear
reason for them to come only from the renormalizable chiral superpotential.
They may also originate from nonrenormalizable terms, which can be functions 
of both left and right chiral superfields.  Such nonholomorphic terms allow 
new supersymmetry-breaking soft terms \cite{hr} in the Lagrangian.  The most 
general set of nonholomorphic soft terms conserving R-parity is: 
\begin{equation}
{\cal L}_{soft}^{NH}=
- N^e_{ij} H_2^{a \ast} \tilde{L}_i^a \tilde{e}_j^c
- N^d_{ij} H_2^{a \ast} \tilde{Q}_i^a \tilde{d}_j^c
- N^u_{ij} H_1^{a \ast} \tilde{Q}_i^a \tilde{u}_j^c \, + \, h.c.
\label{lsoftNH}\,.
\end{equation}
Similarly, nonholomorphic soft terms breaking R-parity are:
\begin{eqnarray}
{\cal L}_{soft}^{NH R \! \! \! /}&=&  
-{N}^{\prime {\tiny B}}_i H_1^{a \ast} \tilde{L}_i^a
- {N}^{\prime e}_{i} H_2^{a \ast} H_1^a \tilde{e}_i^c
- {N}^{\prime u}_{ijk} \tilde{L}_i^{a \ast} \tilde{Q}_j^a \tilde{u}_k^c
\nonumber\\[1.5mm] 
&& 
- {N}^{\prime {\tiny S}}_{ijk} \tilde{u}_i^c \tilde{e}_j^c \tilde{d}_k^{c \ast} 
- {N}^{\prime d}_{ijk} \varepsilon_{ab} \tilde{Q}_i^{a} \tilde{Q}_j^b 
\tilde{d}_k^{c \ast} \, + \, h.c.
\label{lsoftNHRviol}
\end{eqnarray}
In our convention, all ``N" constants are for nonholomorphic terms.
Since we assume lepton-number violation but baryon-number 
conservation, it implies 
$\lambda^{\prime \prime }_{ijk}=A^{\prime {\tiny S}}_{ijk}=
N^{\prime d}_{ijk}=0$. 

\section{Bilinear R-Parity Violation \label{Bilinear section}}

Before coming to the explicit model in the next section, we first look at the 
possibility of generating a lepton asymmetry from bilinear R-parity violating 
terms. This case is a good illustration of the general problems involved. 
One immediate consequence of the violation of lepton number through the 
bilinear terms is the mixing of the neutrinos with the neutralinos.  In the 
MSSM there are four neutralinos, the U(1) gaugino ($\tilde B$), the SU(2) 
gaugino ($\tilde W_3$), and the two Higgsinos ($\tilde h_1^0$, $\tilde h_2^0$).
When lepton number is violated through the R-parity violating terms, it is 
possible to assign zero lepton number to what we usually regard as lepton 
superfields \cite{maroy,cf}.  Suppose only the $\tau$ neutrino mixes with 
the neutralinos, then both $\tau$ and $\nu_\tau$ may be assigned an effective 
vanishing lepton number while the other leptons remain leptons.  However, in 
the general three-family case, all three neutrinos may mix with the 
neutralinos, and the scalar partners of the neutrinos ($\tilde \nu_i$) may 
all acquire nonzero vacuum expectation values (VEVs).

In the most general case the neutralino mass matrix with all the
seven fields in the basis $\left[ ~\tilde B,~~ \tilde W_3,
~~ \tilde h_1^0, ~~\tilde h_2^0, ~~\nu_1, ~~ \nu_2, ~~\nu_3~ \right]$
is given by
\begin{equation}
{\cal M} = \left[ \begin{array} {c@{\quad}c@{\quad}c@{\quad}c@{
\quad}c@{\quad}c@{\quad}c} 
M_1 & 0 & -s r_Z v_1  & s r_Z v_2 & 
-s r_Z v_{\nu_1} & -s r_Z v_{\nu_2} & 
-s r_Z v_{\nu_3}\\ 
0 & M_2 & c r_Z v_1  & -c r_Z v_2 & 
c r_Z v_{\nu_1} & c r_Z v_{\nu_2}& 
c r_Z v_{\nu_3}\\ 
-s r_Z v_1 & c r_Z v_1 & 0 & -\mu & 0 &0 &0\\ 
s r_Z v_2 & -c r_Z v_2 & -\mu & 0 & -\mu_1 & 
-\mu_2 & -\mu_3\\ 
-s r_Z v_{\nu_1} & c r_Z v_{\nu_1} & 0 & -\mu_1 & 0 &0 &0\\ 
-s r_Z v_{\nu_2} & c r_Z v_{\nu_2} & 0 & -\mu_2 & 0 &0 &0\\ 
-s r_Z v_{\nu_3} & c r_Z v_{\nu_3} & 0 & -\mu_3 & 0 &0 &0 
\end{array} \right],
\end{equation}
where $s = \sin \theta_W$, $c = \cos \theta_W$, $r_Z=M_Z/v$, and $v_1$, 
$v_2$, $v_{\nu_i}$ are the VEVs of $h_1^0$, $h_2^0$, and $\tilde \nu_i$ 
respectively, with $v_1^2 +v_2^2 +v_{\nu}^2 =v^2 \simeq$ (246 GeV)$^2$ and 
$v_\nu^2= v_{\nu_1}^2 + v_{\nu_2}^2 +v_{\nu_3}^2$.  We also define 
$\tan \beta = v_2 / (v_1^2 + v_\nu^2)^{1/2}$.

To understand the structure of the above $7 \times 7$ mass matrix, let us 
assume that $\mu$ is the dominant term, then $\tilde h^0_{1,2}$ form 
a heavy Dirac particle of mass $\mu$ which mixes very little  with the other
physical fields.  Removing these heavy fields will then give us the reduced 
$5 \times 5$ matrix in the basis ($\tilde B, \tilde W_3, \nu_1, \nu_2, \nu_3$):
\begin{equation}
{\cal M} = \left[ \begin{array} {c@{\quad}c@{\quad}c@{\quad}c@{\quad}c} 
M_1 - s^2 \delta & sc \delta & -s \epsilon_1 & -s \epsilon_2 & -s \epsilon_3\\ 
sc \delta & M_2 - c^2 \delta &  c \epsilon_1 &  c \epsilon_2 &  c \epsilon_3 \\ 
-s \epsilon_1 & c \epsilon_1 & 0 & 0 & 0\\
-s \epsilon_2 & c \epsilon_2 & 0 & 0 & 0\\
-s \epsilon_3 & c \epsilon_3 & 0 & 0 & 0
\end{array} \right],
\end{equation}
where 
\begin{eqnarray}
\delta &=& 2 M_Z^2 \frac{v_1 v_2}{v^2} \frac{1}{ \mu} = 
\frac{M_Z^2 \sin 2 \beta}{\mu} 
\sqrt{1-\frac{v_{\nu}^2}{v^2 \cos^2 \beta}}, \\ 
\epsilon_i &=& \frac{M_Z}{v} \Big( v_{\nu_i} -  
\frac{\mu_i}{\mu} v_1 \Big).  
\end{eqnarray}
{}From the above, only the combination $\nu_l \equiv (\epsilon_1\nu_1 +
\epsilon_2 \nu_2 + \epsilon_3 \nu_3)/\epsilon$, with 
$\epsilon^2=\epsilon_1^2 + \epsilon_2^2 + \epsilon_3^2$, mixes with the 
gauginos. This state will have an effective vanishing lepton number
and the other two orthogonal combinations decouple from the neutralino
mass matrix. In this case, only the eigenstate
\begin{equation}
\nu'_l = \nu_l + {s \epsilon \over M_1} \tilde B - {c \epsilon \over 
M_2} \tilde W_3,
\end{equation}
gets a seesaw mass, i.e.
\begin{equation}
m_{\nu'_l} = - \epsilon^2 \left( {s^2 \over M_1} + {c^2 \over M_2} \right),
\label{massnu}
\end{equation}
whereas the other two neutrinos remain massless.  They may get masses
through one-loop radiative corrections from the usual trilinear R-parity 
violating terms which we have not yet considered.

The two gauginos mix with the neutrino $\nu_l $ and form mass eigenstates 
given by
\begin{eqnarray}
\tilde B' &=& \tilde B + {sc \delta \over M_1 - M_2} \tilde W_3 - {s \epsilon 
\over M_1} \nu_l, \\ \tilde W'_3 &=& \tilde W_3 - {sc \delta \over M_1 - 
M_2} \tilde B + {c \epsilon \over M_2} \nu_l. \label{W3state}
\end{eqnarray}
The physical states $\tilde B'$ and $\tilde W'_3$ now contain $\nu_l$. 
This gives the main feature of R-parity violation, which is the decay
of the lightest neutralino.  By virtue of their $\nu_l$ components, both 
neutralinos will now decay into a lepton or an antilepton and a weak gauge 
boson, such as $\tilde W'_3 \to l^- W^+$ and $l^+ W^-$, thus violating 
lepton number.   Since the mixing of the neutralinos may also have $CP$ 
violation through the complex gaugino masses (thus making $\delta$ complex), 
a lepton asymmetry may be generated from these decays.  However, the amount of 
asymmetry thus generated is several orders of magnitude too small because 
it has to be much less than $(\epsilon/M_{1,2})^2$, which is of order 
$m_{\nu'_l}/M_{1,2}$, i.e. $< 5 \times 10^{-13}$ if $m_{\nu'_l} < 0.05$ eV 
and $M_{1,2} > 100$ GeV.  In addition, the out-of-equilibrium condition on 
the decay width of the lightest neutralino imposes an upper bound on 
$(\epsilon / M_{1,2})^2$ which is independent of $m_{\nu'_l}$, and that 
also results in an asymmetry very much less than $10^{-10}$.

We now consider the R-parity violating trilinear couplings, i.e. $\lambda$ 
and $\lambda'$ of Eq.~(8).  Since the particles involved should have masses 
at most equal to the supersymmetry breaking scale, i.e. a few TeV, their 
$L$ violation together with the $B + L$ violation by sphalerons \cite{sph} 
would erase any primordial $B$ or $L$ asymmetry of the Universe \cite{erase}. 
To avoid such a possibility, we may reduce $\lambda$ and $\lambda'$ to less 
than about $10^{-7}$, but a typical minimum value of $10^{-4}$ is required 
for realistic neutrino masses in one-loop order \cite{numass}.  Hence it 
appears that the MSSM with R-parity violation is not only unsuitable for 
leptogenesis, it is also a destroyer of any lepton or baryon asymmetry 
which may have been created by some other means before the electroweak phase 
transition.
 
\section{Leptogenesis from Neutralino Decay \label{General mechanism}}
%

{}From the discussion of the previous section we observe that for a 
leptogenesis mechanism to be successful in the MSSM with R-parity violation, 
two requirements have to be fulfilled.  First we must use lepton-number 
violating terms which are not constrained by neutrino masses.  Second we must 
satisfy the out-of-equilibrium condition for the lightest neutralino in such 
a way that the asymmetry is not automatically suppressed.  More explicitly, 
we will consider the possibility that the heavier neutralino does not 
satisfy the out-of-equilibrium condition and decays very quickly, but the 
lighter neutralino decays very slowly and satisfies the out-of-equilibrium 
condition.  Since the asymmetry comes from the interference of the one-loop 
$CP$ violating contribution of the heavier neutralino, it is then 
unsuppressed.  We will demonstrate explicitly in the following how this 
scenario may be realized.

We assume first that $M_1 > M_2$, so that the bino $\tilde B$ is heavier 
than the wino $\tilde W_3$.  While the former couples to both $\bar 
e_{i_{\tiny L}} \tilde e_{i_{\tiny L}}$ and $\bar e^{c}_i \tilde e^{c}_i$, 
the latter couples only to $\bar e_{i_{\tiny L}} \tilde e_{i_{\tiny L}}$, 
because the $e^c_i$ are singlets under $SU(2)_L$.  Since R-parity is 
violated, one combination of the $\tilde e_{i_{\tiny L}}$ and another of the 
$\tilde e^{c}_{i}$ mix with the charged Higgs boson of the supersymmetric 
standard model: $h^\pm = h_2^\pm \cos \beta + h_1^\pm \sin \beta$. 
Let us denote them by $\tilde l_L$ and $\tilde l^c$ respectively.  Their 
corresponding leptons are of course $l_L$ and $l^c$.  Hence both $\tilde B'$ 
and $\tilde W'_3$ may decay into $l^\mp h^\pm$.  

We assume next that the $\tilde l_L$ mixing with $h^-$ is negligible, so that 
the only relevant coupling is that of $\tilde B$ to $\bar l^c h^+$.  Hence 
$\tilde W'_3$ decay (into $l^\mp h^\pm$) is suppressed because it may only 
do so through the small component of $\tilde B$ that it contains, assuming 
of course that all charged sleptons are heavier than $\tilde B$ or $\tilde 
W_3$.

With this choice that the heavier neutralino $\tilde B'$ decays quickly 
and the lighter neutralino $\tilde W'_3$ decays much more slowly, we now 
envisage the following leptogenesis scenario.  At temperatures well above $T = 
M_{SUSY}$, there are fast lepton-number and R-parity violating interactions, 
which will wash out any $L$ or $B$ asymmetry of the Universe in the presence 
of sphalerons.  This will be the case even at temperatures around $M_1$, 
when $\tilde W'_3$ interactions violate $L_i$ as well as $B-3L_i$ for $i 
= e, \mu, \tau$ \cite{erase}.  We assume here that all other supersymmetric
particles are heavier than the neutralinos, so that at temperatures
below $M_1$ we need only consider the interactions of $\tilde B'$ and 
$\tilde W'_3$.  In Figure 1 we show the lepton-number violating processes 
(a) $\tilde B' \leftrightarrow l_R^\pm h^\mp$, where we have adopted the 
more conventional notation of an outgoing $l_R$ in place of an incoming $l^c$.
These processes are certainly still fast and there can be no $L$ asymmetry. 
At temperatures far below the mass of the heavier neutralino, the $\tilde B'$ 
interactions are suppressed and we need only consider those of $\tilde W'_3$. 
With our assumptions, the lepton-number violating processes (b) $\tilde W'_3 
\leftrightarrow l_R^\pm h^\mp$ are slow and will satisfy the 
out-of-equilibrium condition for generating a lepton asymmetry of the 
Universe.  Specifically, it comes from the interference of this tree-level 
diagram with the one-loop (c) self-energy and (d) vertex diagrams.   Since 
the unsuppressed lepton-number violating couplings of $\tilde B'$ are 
involved, a realistic lepton asymmetry may be generated.  It is then 
converted by the still active sphalerons into the present observed baryon 
asymmetry of the Universe.  In this scenario the mass of $\tilde W'_3$ also 
has to be small enough so that the scattering processes mediated by the 
heavier $\tilde B'$ are negligible at temperature below $M_2$ when the 
asymmetry is produced. 

We start with the well-known interaction of $\tilde B$ with $l$ and 
$\tilde l_R$ given by \cite{hk}
\begin{equation}
-{e \sqrt 2 \over \cos \theta_W} \left[ \bar l \left( {1 - \gamma_5 \over 
2} \right) \tilde B \tilde l_R + H.c. \right].
\label{interac1}
\end{equation}
We then allow $\tilde l_R$ to mix with $h^-$, and $\tilde B$ to mix with 
$\tilde W_3$, so that the interaction of the physical state $\tilde W'_3$ 
of Eq.~(\ref{W3state}) with $l$ and $h^\pm$ is given by
\begin{equation}
\left( {sc\xi \delta r \over M_1 - M_2} \right) 
\left( {e \sqrt 2 \over \cos \theta_W} 
\right) \left[ \bar l \left( {1 - \gamma_5 \over 2} \right) \tilde 
W'_3 h^- + H.c. \right],
\label{interac2}
\end{equation}
where $\xi$ represents the $\tilde l_R - h^-$ mixing and is assumed real, 
but the parameter $\delta$ of Eq.~(15) is complex.  We have also inserted 
a correction factor $r=(1+M_2 / \mu \sin 2 \beta)/(1-M_2^2 / \mu^2)$ for 
finite values of $M_2 / \mu$.  The origin of a nontrivial $CP$ phase in the 
above is from the $2 \times 2$ Majorana mass matrix spanning $\tilde B$ and 
$\tilde W_3$, with complex $M_1$ and $M_2$.  It is independent of the phase 
of $\mu$ and contributes negligibly to the neutron electric dipole moment 
because the magnitude of $\delta$ is very small.  [Note that the usual 
assumption of $CP$ violation in supersymmetric models is that $M_1$ and 
$M_2$ have a common phase, in which case the phase of $\delta$ would be 
equal to the phase of $\mu$.]

The decay width of the bino is then 
\begin{equation}
\Gamma_{\tilde{B}'} = 
\Gamma ({\tilde{B}'} \to {l^{+}} {h^{-}})+
\Gamma ({\tilde{B}'} \to {l^{-}} {h^{+}})=
\frac{1}{4 \pi} \xi^2 
\frac{e^2}{c^2}
 {(M_{\tilde{B}'}^2 - m_h^2)^2 \over 
M_{\tilde{B}'}^3},
\end{equation}
while that of the wino is
\begin{equation}
\Gamma_{\tilde{W}'_3} = 
\Gamma ({\tilde{W}'_3} \to {l^{+}} {h^{-}})+
\Gamma ({\tilde{W}'_3} \to {l^{-}} {h^{+}})
=\frac{1}{4 \pi} \xi^2 
\left( { e s |\delta| r \over M_1 - M_2} \right)^2 
{(M^2_{\tilde{W}'_3} - m_h^2)^2 \over 
M_{\tilde{W}'_3}^3}.\label{decW}
\end{equation}

{}Using Eqs.~(\ref{interac1}) and (\ref{interac2}), we calculate the 
interference between the tree-level and self-energy + vertex diagrams of 
Figure 1 and obtain the following asymmetry from the decay of $\tilde W'_3$:
\begin{equation}
\epsilon = \frac{\Gamma ({\tilde{W}'_3} \to {l^{+}} {h^{-}})-
\Gamma ({\tilde{W}'_3} \to {l^{-}} {h^{+}})}{\Gamma_{\Wp}}
= {\alpha \xi^2 \over 2 \cos^2 \theta_W} 
{Im \delta^2 \over 
|\delta|^2} \left( 1 - {m_h^2 \over \MWp^2} \right)^2 {x^{1/2} g(x) \over 
(1-x)}, \label{dasym}
\end{equation}
where $x= M_{\tilde{W}'_3}^2/M_{\tilde{B}'}^2$ and
\begin{equation}
g(x) = 1 + {2(1-x) \over x} \left[ \left( {1+x \over x} \right) \ln (1+x) 
-1 \right].
\end{equation}
If the $\tilde W'_3$ interactions satisfy the out-of-equilibrium condition, 
then a lepton asymmetry may be generated from the above decay asymmetry. 
Note that in the above expression for $\epsilon$, the parameter $\delta$ 
appears only in the combination $ {\rm Im} \delta^2 / |\delta|^2 $, 
which may be of order one.  If the absolute value of $\delta$ is small, it 
slows down the decay rate of $\tilde W'_3$ and a departure from equilibrium 
may be achieved without affecting the amount of decay asymmetry generated 
in the process.  

At the time this lepton asymmetry is generated, if the sphaleron interactions 
\cite{sph} are still in equilibrium, they will convert it into a baryon 
asymmetry of the Universe \cite{ht}.  If the electroweak phase transition 
is strongly first-order, the sphaleron interactions freeze out 
at the critical temperature.   Lattice simulations suggest that for a 
Higgs mass of around $m_H \approx 70$ GeV, the critical temperature is
around $T_c \approx 150$ GeV \cite{latt}.  Higher values of $m_H$ will 
increase the critical temperature, but the increase is slower than 
linear.  For example, for $m_H \approx 150$ GeV, the critical temperature 
could go up to $T_c \approx 250$ GeV. 

For a second-order or weakly first-order phase transition\footnote{
Note that we do not require the electroweak phase transition to be 
first-order for satisfying the out-of-equilibrium condition. See for example 
Ref.~\cite{riotto}.}, the sphaleron 
interactions freeze out at a temperature lower than the critical temperature. 
After the electroweak phase transition ($T < T_c$), the sphaleron 
transition rate is given by \cite{spra}
\begin{equation}
\Gamma_{sph} (T) = (2.2 \times 10^4 ~ \kappa) ~ {[ 2 ~ M_W (T) ]^7
\over [ 4 \pi ~ \alpha_W T ]^3 } ~ e^{- E_{sph(T)} /T },
\end{equation}
where 
\begin{equation}
M_W(T) = {1 \over 2} g_2 \langle v(T) \rangle
= {1 \over 2} g_2 \langle v(T=0) \rangle ~ \left( 1 - {T^2 \over 
T_c^2} \right)^{1/2},
\end{equation}
and the free energy of the sphaleron is $E_{sph}(T) \approx ( 2 ~ M_W(T) /
\alpha(W) ) ~ B( m_h / M_W)$, with $B(0) = 1.52$, $B(\infty) = 2.72$
and $\kappa = e^{-3.6}$ \cite{moore}.  In this case, the sphalerons 
freeze out at a temperature $T_{out}$ which is the temperature at which 
their interaction strength equals the expansion rate of the universe,
\begin{equation}
\Gamma_{sph}(T_{out}) = H(T_{out}) = 1.7 \sqrt{g_*} {T^2_{out}
\over M_{Pl} } .
\end{equation}
For a critical temperature of about $T_c \sim 250$ GeV, the freeze-out 
temperature comes out to be around $T_{out} \sim 200$ GeV. 

These discussions indicate that as long as the lepton asymmetry is 
generated at a temperature above, say 200 GeV, it will be converted to 
a baryon asymmetry of the Universe.   Since the sphaleron interactions 
grow exponentially fast, they can convert a lepton asymmetry to a baryon 
asymmetry [a $(B-L)$ asymmetry to be precise] by the time the temperature 
drops by only a few GeV.  In the next section we discuss how the decay 
asymmetry of the neutralinos becomes a lepton asymmetry of the Universe.

%
\section{Boltzmann Equations \label{Boltzmann}}

We now solve the Boltzmann equations \cite{kt} to estimate the amount of
lepton asymmetry created after the decays of the neutralinos.
When the decay of the $\tilde W'_3$ satisfies the out-of-equilibrium 
condition, i.e. when the decay rate is slower than the expansion rate
of the Universe, the generated asymmetry is of the order of the decay 
asymmetry given in Eq.~(\ref{dasym}). This argument could replace the 
details of solving the Boltzmann equations for an order-of-magnitude
estimate of the asymmetry in many scenarios.  However, in the present 
case there are other constraints and depleting factors, and we need 
to solve the Boltzmann equations explicitly for a reliable estimate.

If the $\tilde{W}'_3$ decay rate is much less than the expansion rate 
of the Universe, the generated lepton asymmetry is the same as the decay 
asymmetry.  In other words, the out-of-equilibrium condition reads 
\begin{equation}
K_{\Wp} = {\Gamma_{\tilde{W}'_3} \over
H(M_{\tilde{W}'_3})} \ll 1 , 
\end{equation}
where $H(T)$ is the Hubble constant at the temperature $T$ and is given by
\begin{equation}
H(T)=\sqrt{\frac{4 \pi^3 g_\ast}{45}} \frac{T^2}{M_{Planck}},\label{hubble}
\end{equation}
with $g_\ast$ the number of massless degrees of freedom which we take equal 
to $106.75$ and $M_{Pl} \sim 10^{18}$ GeV is the Planck scale.  If this 
condition is satisfied, the lepton asymmetry is given by $n_L = n_l - 
n_{\bar l}  \sim \epsilon/g_*$.  But in practice, when $K \ll 1$, there is 
no time for the asymmetry to grow to its maximum value before the sphaleron 
transitions are over.  So we need to study the case $K \sim 1$. Furthermore,
a reasonable amount of asymmetry cannot be obtained 
unless the inverse decay and the scattering from bino exchange have rates 
lower than the expansion rate of the Universe.  All these effects result 
in the further diminution of the lepton asymmetry and we need to solve the 
Boltzmann equations to take care of them properly.

At temperatures $T<M_2$, the decays of $\tilde{W}'_3$ given in Eq.~(\ref{decW})
start generating an asymmetry.  At this time there are important damping 
contributions coming from the inverse decays of $\tilde{W}'_3$ and 
$\tilde{B}'$ as well as the scattering processes $l^{\pm} + h^\mp \rightarrow 
\tilde{B}' \rightarrow l^\mp  + h^\pm$.  As we will see, the last two 
processes are especially important because $\tilde{B}'$ tends to remain 
in equilibrium and its presence washes out the created lepton asymmetry from 
the $\tilde W'_3$ decays.  The reason is that its interactions are strong 
enough so that the Boltzmann exponential suppression of its number density may 
not be sufficient to compensate its large inverse decay and scattering cross 
sections.  The effect of the scattering $ l^\pm + h^\mp \rightarrow 
\tilde{W}'_3 \rightarrow l^\mp + h^\pm$ is on the other hand negligible 
because it is suppressed by a factor of $[(s c \delta r)/(M_1-M_2)]^2 $ 
with respect to the scattering $l^{\pm} + h^\mp \rightarrow \tilde{B}' 
\rightarrow l^\mp  + h^\pm$.  Neglecting this term and defining the variable 
$z \equiv M_{\tilde{W}_3}/T$, the Boltzmann equations are then:
\begin{eqnarray}
\frac{dX_{\tilde{W}'_3}}{d z}&=&
-\gamma^{eq}_{\tilde{W}'_3}   \frac{z}{sH(M_{\tilde{W}'_3})}  
\Big(\frac{X_{\tilde{W}'_3}}{X^{eq}_{\tilde{W}'_3}} - 1 \Big),\label{boltz1}\\
\frac{d X_L}{d z}&=&
\gamma^{eq}_{\tilde{W}'_3}   \frac{z}{sH(M_{\tilde{W}'_3})} 
\Big[\varepsilon \Big(\frac{X_{\tilde{W}'_3}}{X^{eq}_{\tilde{W}'_3}} - 1 \Big)
- \frac{1}{2} \frac{ X_L}{X_\gamma} \Big]\nonumber\\
&&-  \frac{z}{sH(M_{\tilde{B}'})} \Big(\frac{M_{\tilde{B}'}}
{M_{\tilde{W}'_3}} \Big)^2 
\Big[ \gamma^{eq}_{\tilde{B}'} \frac{1}{2} \frac{X_L}{X_\gamma} 
+2 \frac{X_L}{X_\gamma} \gamma^{eq}_{scatt.} \Big],\label{boltz2}
\end{eqnarray}    
where we have defined the number densities per comoving volume 
$X_i = n_i / s$ in terms of the number densities of particles ``i" and 
\begin{equation}
s=g_\ast \frac{2 \pi^2}{45} T^{3}\label{entropy}
\end{equation}
is the entropy density.  The equilibrium distributions of the number 
densities are given by the Maxwell-Boltzmann statistics:
\begin{eqnarray}
n_{\tilde{W}'_3}&=&g_{\tilde{W}'_3}\frac{M_{\tilde{W}'_3}^2}{2 \pi^2}
T K_2(M_{\tilde{W}'_3}/T),    \\
n_{\tilde{B}'}&=&g_{\tilde{B}'}\frac{M_{\tilde{B}'}^2}{2 \pi^2}
T K_2(M_{\tilde{B}'}/T),    \\
n_\gamma&=&\frac{g_{\gamma} T^3}{\pi^2},
\end{eqnarray}
where $g_{\tilde{W}'_3}=1$, $g_{\tilde{B}'}=1$, and $g_\gamma =2$ are the 
numbers of degrees of freedom of $\tilde{W}'_3$, $\tilde{B}'$, and the 
photon respectively. 

The quantities $\gamma^{eq}_{\Wp}$ and $\gamma^{eq}_{\Bp}$ are the 
reaction densities for the decays and inverse decays of $\Wp$ and $\Bp$:
\begin{eqnarray}
\gamma^{eq}_{\tilde{W}'_3}&=&n^{eq}_{\tilde{W}'_3} 
\frac{K_1(M_{\tilde{W}'_3}/T)}{K_2(M_{\tilde{W}'_3}/T)} 
\Gamma_{\tilde{W}'_3}, \\
\gamma^{eq}_{\tilde{B}'}&=&n^{eq}_{\tilde{B}'} 
\frac{K_1(M_{\tilde{B}'}/T)}{K_2(M_{\tilde{B}'}/T)} 
\Gamma_{\tilde{B}'},
\end{eqnarray}
$K_1$ and $K_2$ being the usual modified Bessel functions.  The reaction 
density for the scattering is given by
\begin{equation}
\gamma^{eq}_{scatt.}= \frac{T}{64 \pi^4} \int_{(m_h+m_l)^2}^{\infty}
ds\, \hat{\sigma}_{\tilde{B}'}(s)\, \sqrt{s} \, K_1(\sqrt{s}/T),
\label{scattint}
\end{equation}
where $\hat{\sigma}_{\tilde{B}'}$ is the reduced cross section and is given 
by $2 [s-(m_h+m_l)^2][s-(m_h-m_l)^2] \sigma_{\tilde{B}'}/s \, \sim \, 2s 
\sigma_{\tilde{B}'}$.   The cross section $\sigma_{\tilde{B}'}$ does not 
contain the contribution of the on-mass-shell bino (which is already 
taken into account in the decay and inverse decay terms).  This is achieved 
by replacing the usual propagator $1/(s-m^2+i\Gamma m)$ with the 
off-mass-shell propagator \cite{Luty,Plum}:
\begin{equation}
D^{-1}_s=\frac{s-m^2 }{(s-m^2)^2 +\Gamma^2 m^2}.
\end{equation}
The cross section $\sigma_{\tilde{B}'}$ contains the s- and t-channel 
contributions together with their interference terms and is given by
\begin{eqnarray}
&&\sigma_{\tilde{B}'}\equiv
\sigma (l^{\pm} h^{\mp} \rightarrow {\tilde{B}'} \rightarrow 
l^{\mp} h^{\pm})\,\,=\,\, \nonumber\\
&& \frac{1}{8 \pi s^2} \Big(\frac{e^2 \xi^2}
{cos^2 \theta_W}\Big)^2 m_{\tilde{B}'}^2  
\, \Big[ \frac{s^2}{D_s^2} + \frac{4 s}{D_s} +\frac{2 s}{M_{\tilde{B}'}^2}
-\Big( 2 + 4\frac{s + M_{\tilde{B}'}^2}{D_s} \Big)
\ln\Big( 1+\frac{s}{M_{\tilde{B}'}^2} \Big) \Big]\,
\end{eqnarray}
In Eq.~(\ref{scattint}) the integral is dominated by the s-channel 
contribution in the resonance region and to a very good approximation, 
$\gamma^{eq}_{scatt.}$ reduces to
\begin{equation}
\gamma^{eq}_{scatt.}= \frac{T}{512 \pi^4} \frac{M_{\tilde{B}'}^4}
{\Gamma_{\tilde{B}'} }
K_1(M_{\tilde{B}'}/T)  \frac{e^4 \xi^4}{cos^4 \theta_W}.
\label{scattres}
\end{equation}
{}From Eqs.~(\ref{entropy}) to (\ref{scattres}), we find the Boltzmann 
equations, i.e. (\ref{boltz1}) and (\ref{boltz2}), to be given by
\begin{eqnarray}
\frac{dX_{\tilde{W}'_3}}{d z}&=&
- z K_{\tilde{W}'_3} \frac{K_1(z)}{K_2(z)}   
\Big({X_{\tilde{W}'_3}}-{X^{eq}_{\tilde{W}'_3}}  \Big),\nonumber\\
\frac{d X_L}{d z}&=&
z K_{\tilde{W}'_3}   \frac{K_1(z)}{K_2(z)} 
\Big[\varepsilon ({X_{\tilde{W}'_3}}-{X^{eq}_{\tilde{W}'_3}} )
- \frac{1}{2}   \frac{X_{\tilde{W}_3}}{X_\gamma} X_L \Big]\nonumber\\
&&- z  \Big(\frac{M_{\tilde{B}'}}
{M_{\tilde{W}'_3}} \Big)^2 
K_{\tilde{B}'} \Big[\frac{1}{2} \frac{K_1(z M_{\tilde{B}'}/M_{\tilde{W}'_3})}
{K_2(z M_{\tilde{B}'}/M_{\tilde{W}'_3})}
\frac{X_{\tilde{B}'}}{X_\gamma} X_L 
+2 \frac{X_L}{X_\gamma} 
\frac{\gamma^{eq}_{scatt.}}{s 
\Gamma_{\tilde{B}'} } \Big],\label{boltzmann}
\end{eqnarray}    
with
\begin{equation}
K_{\tilde{B}'}=\Gamma_{\tilde{B}'} /
H(M_{\tilde{B}'}),
\end{equation}
which gives the strength of lepton-number violation in the decays of $\Bp$. 

If we now ignore the inverse decay and scattering processes, we can simplify 
the problem by requiring the out-of-equilibrium condition to be 
\begin{equation}
 K_{\tilde{W}'_3} < 1.
\label{oecW}
\end{equation}
With the terms proportional to $K_{\tilde{B}'}$ this condition is necessary 
but not sufficient.  In the present scenario for a large asymmetry 
we also require $K_{\tilde{B}'} > 1$. Indeed, $K_{\tilde{B}'}$ is larger
than $K_{\tilde{W}'_3}$ by a factor $R_K=K_{\tilde{B}'}/K_{\tilde{W}'_3}
 \sim [(s c \delta r)/ (M_1-M_2)]^{-2} (M_{\tilde{W}'_3}/ M_{\tilde{B}'})
$, which is larger than one by several orders of magnitude. Therefore 
${\tilde{B}'}$ remains in equilibrium and the $\tilde{B}'$ damping terms,
due to its inverse decay and scattering, dominate over the $\tilde{W}'_3$ 
inverse decay damping term as long as the Boltzmann suppression factor in 
the $\tilde{B}'$ equilibrium distribution has not compensated the large 
value of $R_K$.  For example with the set of parameters 
$M_{\tilde {W}'_3}=2$ TeV, $\MBp=3$ TeV, 
$\sin 2\beta=0.5$, $\xi=2 \times 10^{-3}$, $\mu= 5$ TeV used
in Ref.~\cite{hms}, we obtain 
$K_{\tilde{W}'_3}=0.63$ and $K_{\tilde{B}'}=7.8 \times 10^{5}$, and the 
$\Bp$ damping terms dominate over those of $\Wp$ as long as the temperature 
is above $\sim 65$ GeV (the former differs from the latter by a factor 
$\sim R_K (\MBp / \MWp)^3 e^{-(\MBp - \MWp)/T}$).  In this case the inverse 
decay and scattering of $\Bp$ cause a considerable wash-out of the asymmetry 
because $\Wp$ has mostly decayed away already at temperatures well above 
$\sim$~65 GeV.  This is illustrated in Figure 2 showing the effects of 
various terms in the Boltzmann equations.

To avoid this wash-out, the value of $\MBp$ has to be larger in order that 
the $\Bp$ number density is further suppressed at temperatures below $\MWp$ 
when the asymmetry is produced.  Varying the parameters of these $\Bp$ 
damping terms, it appears difficult to induce a sufficiently large asymmetry 
of order $10^{-10}$ for $\MBp$ below 4 TeV.  Two typical situations for 
which a large asymmetry is produced are for example: 
\begin{eqnarray}
&& \MBp = 6 ~{\rm TeV}, ~~~ \MWp = 3.5 ~{\rm TeV}, ~~~ \xi = 5 \times 10^{-3},
\nonumber 
\\ && \mu = 10 ~{\rm TeV}, ~~~ \sin 2 \beta = 0.10, ~~~ m_h = 200 ~{\rm GeV},
\label{set1}
\end{eqnarray}
and
\begin{eqnarray}
&& \MBp = 5 ~{\rm TeV}, ~~~ \MWp = 2 ~{\rm TeV}, ~~~ \xi = 5 \times 10^{-3},
\nonumber 
\\ && \mu = 7.5 ~{\rm TeV}, ~~~ \sin 2 \beta = 0.05, ~~~ m_h = 200 ~{\rm GeV},
\label{set2}
\end{eqnarray}
for which we have $K_{\Wp} =0.02$, $K'_{\Bp} =2.4 \times 10^{6}$ and
$K_{\Wp}=0.02$, $K'_{\Bp}=2.9 \times 10^{6}$ respectively.  At $T=M_Z$ 
the leptonic asymmetry produced is $X_L=1.0 \times 10^{-10}$ with the 
parameters of Eqs.~(\ref{set1}) and $X_L=1.2 \times 10^{-10}$ with those 
of Eq.~(\ref{set2}).  Figures 3 and 4 show the evolution of the asymmetry 
in these two cases.  As can be seen from these figures, the damping effects 
of the inverse decay of $\Wp$ and of the scattering are small\footnote{
Note that in the Boltzmann equations we neglected the damping contributions 
of the scatterings $l^\pm l^\pm \rightarrow h^\pm h^\pm$ mediated by a 
neutralino in the t-channel. Their effect
is negligible within the ranges of the parameters we consider.}.  The damping 
effect from the inverse decay of $\Bp$ is however not small and reduces the 
asymmetry by a factor of 2 to 4 by washing out all the asymmetry produced 
above $T \sim 300-400$~GeV.

A large asymmetry of order $10^{-10}$ is produced provided $\MBp$ is of the 
order 4 TeV or more.  A low value of $\sin 2\beta$ below $\sim 0.30$ is 
generally necessary.  Values of $\xi$ around $3-5 \times 10^{-3}$, of $\mu$ 
around $5-10$ TeV, and of $\MWp$ from 1 TeV to $2 \MBp /3$ are also preferred.

%
%
\section{Charged Scalar Mass Matrix \label{scalar}}

The mechanism we propose for leptogenesis requires the decay of $\tilde B'$ 
to be fast, while that of $\tilde W_3'$ is very slow.  This is achieved by 
requiring $l_R^\pm h^\mp$ to be the main decay mode and $l_L^\pm h^\mp$ to 
be negligible.  Hence $\tilde l_R$ must mix with $h^-$ readily, so that 
$\tilde B'$ could decay directly, but $\tilde W_3'$ could decay only through 
its small $\tilde B$ component.  We now consider the charged scalar mass 
matrix which determines this mixing.  As shown in the following, our present 
scenario requires one more new ingredient, i.e. the presence of nonholomorphic 
soft terms. 

The value of the $\tau_R-h^+$ mixing parameter $\xi$ is governed by the 
charged scalar mass matrix which follows from the quadratic terms in the 
Lagrangian:
\begin{equation}
{\cal L} \owns - {\Phi}^\dagger {\mathcal{M}}^2_{S^\pm} {\Phi},
\end{equation}
with $\Phi= [ h_1^{- \ast},h_2^+,\tilde e_{i_{\tiny L}}^{- \ast}, \tilde 
e_i^c ]^T$.  In the case where all $N$ constants are put to zero and for 
parameters satisfying the various constraints from the Boltzmann equations 
(see previous section), it is difficult to generate a sufficiently large 
value of the $\tilde l^c-h^-$ mixing parameter $\xi$ (see Appendix A). Thus 
we will ignore the $\mu_i$ terms and the associated vaccuum expectation 
values $v_{\nu_i}$ in the following.  To induce a large $\tilde e^c-h^-$ 
mixing, we introduce the nonholomorphic terms of Eqs.~(\ref{lsoftNH}) and 
(\ref{lsoftNHRviol}). The mass matrix is then given by
\begin{equation}
{\mathcal{M}}^2_{S^{\pm}}=\left(
\begin{array}{cccc}
{\mathcal{M}}^2_{{h_1^-}-{h_1^+}} & 
{\mathcal{M}}^2_{{h_1^-}-{h_2^+}} & 
{\mathcal{M}}^2_{{h_1^-}-{\tilde e_{i_{\tiny L}}}^*} & 
{\mathcal{M}}^2_{{h_1^-}-{\tilde e^c_i}}\\
{\mathcal{M}}^2_{{h_2^-}-{h_1^+}} & 
{\mathcal{M}}^2_{{h_2^-}-{h_2^+}} & 
{\mathcal{M}}^2_{{h_2^-}-{\tilde e_{i_{\tiny L}}}^*} & 
{\mathcal{M}}^2_{{h_2^-}-{\tilde e^c_i}}\\
{\mathcal{M}}^2_{{\tilde e_{j_{\tiny L}}}-{h_1^+}} & 
{\mathcal{M}}^2_{{\tilde e_{j_{\tiny L}}}-{h_2^+}} & 
{\mathcal{M}}^2_{{\tilde e_{j_{\tiny L}}}-\tilde e_{i_{\tiny L}}^*} & 
{\mathcal{M}}^2_{{\tilde e_{j_{\tiny L}}}-{\tilde e^c_i}}\\
{\mathcal{M}}^2_{{\tilde e_{j}^c}-{h_1^+}} & 
{\mathcal{M}}^2_{{\tilde e_{j}^c}-{h_2^+}} & 
{\mathcal{M}}^2_{{\tilde e_{j}^c}-\tilde e_{i_{\tiny L}}^*} & 
{\mathcal{M}}^2_{{\tilde e_{j}^c}-{\tilde e^c_i}}
\end{array} \right)\,,\label{massmatrixdef}
\end{equation}
where
\begin{eqnarray}
{\mathcal{M}}^2_{{h_1^-}-{h_1^+}}&=&
\frac{g^2}{4} v_2^2 
-B \frac{v_2}{v_1},
\\
{\mathcal{M}}^2_{{h_1^-}-{h_2^+}}&=&
 \frac{g^2}{4} v_1 v_2  -B, \\
{\mathcal{M}}^2_{{h_1^-}-{\tilde e_{i_{\tiny L}}^*}}&=&
 N^{\prime {\tiny B}}_i, \\
{\mathcal{M}}^2_{{h_1^-}-{\tilde e_i^c}}&=&
\frac{1}{\sqrt{2}} N^{\prime e}_i v_2, \\
{\mathcal{M}}^2_{{h_2^-}-{h_2^+}}&=&
\frac{g^2}{4} v_1^2 
-B \frac{v_1}{v_2},
\\
{\mathcal{M}}^2_{{h_2^-}-{\tilde e_{i_{\tiny L}}^*}}&=&
-B'_{i}, \\
{\mathcal{M}}^2_{{h_2^-}-{\tilde e_i^c}}&=&
\frac{1}{\sqrt{2}} N^{\prime e}_i v_1, \\
{\mathcal{M}}^2_{{\tilde e_{j_{\tiny L}}}-{\tilde e_{i_{\tiny L}}^*}}&=&
(M^2_L)_{ij} -\frac{1}{8} (g^2-g'^2)(v_1^2-v_2^2) \delta_{ij}
+\frac{1}{2}f^e_{jk} f^e_{ik} v_1^2,  
\\
{\mathcal{M}}^2_{{\tilde e_{j_{\tiny L}}}-{\tilde e_{i}^c}}&=&
\frac{1}{\sqrt{2}}f^e_{ji} \mu v_2 + \frac{1}{\sqrt{2}} (A_e f_e)_{ji} {v_1}
+\frac{1}{\sqrt{2}}N^e_{ji} v_2,
\\
{\mathcal{M}}^2_{{\tilde e_{j}^{c*}}-{\tilde e_{i}^c}}&=&
(M^2_{\tilde{e}^c})_{ji}
-\frac{g'^2}{4} (v_1^2-v_2^2)\delta_{ij}
+\frac{1}{2}f^e_{ki} f^e_{kj} v_1^2. 
\end{eqnarray}

In the following we assume that $N_i^{\prime {\tiny B}}$ and $B^{\prime}_i$ 
are negligible so that they do not induce a large mixing of the left-handed 
charged sleptons with the charged Higgs boson.  We are then left with the 
$N_i^{\prime e}$ and $N^e_{ji}$ terms.  Going to the basis of the physical 
charged Higgs boson $h^+$ and of the Goldstone boson $G^+$, the latter 
decouples and in the basis $[h^+,\tilde e^{- \ast}_{i_{\tiny L}}, \tilde 
e_i^c]$ we obtain the mass matrix
\begin{equation}
{\mathcal{M}}^2_{S^{\pm}}=\left(
\begin{array}{ccc}
m_W^2-2\frac{B}{\sin 2 \beta} & 0 & \frac{1}{\sqrt{2}} N^{\prime e}_i v \\
0  &  
{\mathcal{M}}^2_{{\tilde e_{j_{\tiny L}}}-\tilde e_{i_{\tiny L}}^*} & 
{\mathcal{M}}^2_{{\tilde e_{j_{\tiny L}}}-{\tilde e^c_i}}\\
\frac{1}{\sqrt{2}} N^{\prime e}_i v &
{\mathcal{M}}^2_{{\tilde e_{j}^c}-\tilde e_{i_{\tiny L}}^*} & 
{\mathcal{M}}^2_{{\tilde e_{j}^c}-{\tilde e^c_i}}
\end{array} \right).\label{massmatrix2}
\end{equation}
We observe that only one combination of the charged right-handed sleptons 
mixes with the charged Higgs boson:
\begin{equation}
\tilde{l}^c=\frac{N^{\prime e}_1 \tilde{e}^c_1 + N^{\prime e}_2 \tilde{e}^c_2
+ N^{\prime e}_3 \tilde{e}^c_3}{N^{\prime e}},
\end{equation}
with $(N^{\prime e})^2=(N^{\prime e}_1)^2 + (N^{\prime e}_2)^2 + 
(N^{\prime e}_3)^2$.  In the Lagrangian the mass term which couples the 
charged Higgs boson with the sleptons reduces therefore to the following 
single term:
\begin{equation}
{\cal L} \owns -\frac{1}{\sqrt{2}} v N^{\prime e} h^+ \tilde{l}^c\, + \, h.c.
\end{equation}
In the case of one family, we obtain to a very good approximation:
\begin{eqnarray}
\xi&=&\frac{\frac{1}{\sqrt{2}} N^{\prime e} v}{m^2_{h^+} 
-M^2_{\tilde{e}_{\tiny L}- 
\tilde{e}^\ast_{\tiny L}} - 
\frac{\big( M^2_{\tilde{e}_{\tiny L} - \tilde{e}^c} \big)^2}
{m^2_{h^+}-M^2_{\tilde{e}_{\tiny L}- \tilde{e}^\ast_{\tiny L}}}},\\
\xi'&=& \frac{M^2_{\tilde{e}_{\tiny L}- \tilde{e}^c}}{m^2_{h^+} - 
M^2_{\tilde{e}_{\tiny L}- \tilde{e}^\ast_{\tiny L}}} \, \xi,
\end{eqnarray}
where $\xi'$ is the $h^+ - \tilde e_L$ mixing and
\begin{equation}
m^2_{h^+} = m^2_W - 2 \frac{B}{\sin 2 \beta}.
\end{equation}
As required in section 3, the mixing $\xi'$ has to be much smaller than the 
mixing $\xi$ in order to avoid having the transition $\Wp \rightarrow 
\tilde{e}^{\pm}_L e^\mp_L \rightarrow h^\pm e^\mp_L$. This requires
\begin{equation}
| \xi' | \ < \Big| \xi  \frac{2 s^2 \delta r}{M_1 -M_2} \Big|
\end{equation}
which for the parameters of Eq.~(\ref{set1}) implies that $\xi' < 7 \times 
10^{-5} \, \xi$. Our mechanism requires also that $m_{h^+} < \MWp$ and that 
the mass of any charged slepton is larger than $\MWp$.  We then have
\begin{equation}
\xi \simeq 
\frac{\frac{1}{\sqrt{2}} N^{\prime e} v}{m^2_{h^+} -
M^2_{\tilde{e}_{\tiny L}- \tilde{e}^\ast_{\tiny L}}}, 
\end{equation} 
\begin{equation}
\Big| m_e \mu \tan \beta + A_e m_e + \frac{1}{\sqrt{2}} N_e v_2 \Big| < 
M^2_{\tilde{e}_{\tiny L}- \tilde{e}^\ast_{\tiny L}} \, 
\Big| \frac{2 s^2 \delta r}{M_1 -M_2} \Big|,
\end{equation}
which for the set of parameters given in Eq.~(\ref{set1}) and for 
$M^2_{\tilde{e}_{\tiny L}- \tilde{e}^\ast_{\tiny L}} \sim \MBp^2$ gives 
$N^{\prime e} \sim 1$~TeV and $| m_e \mu \tan \beta + A_e m_e + \frac{1}
{\sqrt{2}} N_e v_2 | < (50\, {\rm GeV})^2$. The latter condition requires 
$m_e < 0.02$ GeV, $|A_e m_e| < (50 \,{\rm GeV})^2$, and $|N_e| < 15$ GeV, 
or a cancellation of the three terms together.  In the case where there is 
no cancellation between these terms the charged slepton which mixs with 
the charged Higgs boson must have predominantly an electron or $\mu$ flavor. 
In the case of a cancellation between these terms, all flavors are possible. 
For other sets of parameters which lead to large asymmetries, these 
numerical bounds can be relaxed easily by a factor of 2 to 4.
In the more general case of three families, similar constraints and 
relations are obtained.

%
%
%
\section{Two-Loop Neutrino Mass}

It is interesting to note that in addition to inducing a lepton asymmetry, 
the nonholomorphic terms $N^{\prime e }_i$ could also generate a neutrino 
mass. Since lepton number is violated at most by one unit in each term, 
the neutrino mass should include at least two lepton-violating vertices in 
a loop diagram.

There exists a one-loop diagram (Fig.~5), which contributes to the 
sneutrino ``Majorana'' mass.  In general, the sneutrinos could have 
diagonal lepton number conserving masses, i.e. $\tilde \nu^* \tilde \nu$.  
But there can be also lepton-number violating mass terms, i.e. $\tilde \nu 
\tilde \nu$ \cite{snu1,snu2,snu3}.  In the present model, the nonholomorphic 
terms give rise to a lepton-number violating sneutrino-antisneutrino mixing 
term, i.e. ${\cal L} \owns - \frac{1}{2} \delta 
m^2_{\tilde \nu_{ij}} \tilde{\nu}_i 
\tilde{\nu}_j + \, h.c.$.  In the case of one family, we get
\begin{equation}
\delta m^2_{\tilde \nu} \sim \frac{1}{8 \pi^2} \frac{\mu^2 \xi^2}{v^2}
 m_l^2.
\label{mnutilde}
\end{equation}
This lepton-number violating sneutrino mass can then induce a Majorana 
neutrino mass \cite{snu2}. 

In the present case the one-loop diagram of Fig.~6 gives a neutrino mass
\begin{eqnarray}
m_\nu &\sim& \frac{1}{32 \pi^2} \frac{e^2}{\sin^2 \theta_W}
\delta m^2_{\tilde \nu}  \MWp
\frac{
 M^2_{\tilde \nu} -\MWp^2- \MWp^2 \ln \Big( 
 M^2_{\tilde \nu}  / \MWp^2 \Big) }{\Big(
 M^2_{\tilde \nu} - \MWp^2 \Big)^2} \nonumber \\
&\sim& \frac{1}{256 \pi^4} \frac{e^2}
{\sin^2 \theta_W} \mu^2
\frac{m_l^2}{v^2} \xi^2 \MWp
\frac{
 M^2_{\tilde \nu} -\MWp^2- \MWp^2 \ln \Big( 
 M^2_{\tilde \nu}  / \MWp^2 \Big) }{\Big(
 M^2_{\tilde \nu} - \MWp^2 \Big)^2}.
\label{mnu}
\end{eqnarray}
In the case where the lepton $l$ which mixes with $h^+$ is essentially 
$\tau$, we get
\begin{eqnarray}
m_\nu&=&\frac{1}{256 \pi^4} 
\frac{e^2}{\sin^2 \theta_W} \mu^2
\frac{m_{\tau}^2}{v^2} \xi^2  \MWp
\frac{
 M^2_{\tilde \nu} -\MWp^2- \MWp^2 \ln \Big( 
 M^2_{\tilde \nu}  / \MWp^2 \Big) }{\Big(
 M^2_{\tilde \nu} - \MWp^2 \Big)^2},
\end{eqnarray}
which has the correct order of magnitude. For example with the 
parameters of Eq.~(\ref{set1}) and taking 
$M_{\tilde \nu} \sim \MBp$  we get $m_{\nu_\tau} \sim 0.1$~eV.  The value 
of $\xi$ we need for having the right order of magnitude for the asymmetry 
is therefore also the one we need to have a neutrino mass, in agreement with 
the present data on atmospheric neutrinos.  In Eq.~(\ref{mnu}), the factor 
$\delta m^2_{\tilde \nu}$ appears because there is GIM 
(Glashow-Iliopoulos-Maiani) suppression from summing over all possible 
neutral slepton eigenstates in the loop.  In Fig.~5  the two-point functions 
of the form $f(m^2,m^{\prime 2},p) \equiv (i/\pi^2) \int d^4 k (k^2-m^2)^{-1} 
((k+p)^2 - m'^2)^{-1}$ have been (roughly) approximated by $\sim 1$ while in 
Fig.~6 the two-point functions have been calculated explicitly (as required 
by the fact that for these diagrams, a GIM suppression mechanism is operative).  This can 
also be understood from another point of view.  Since the diagonal terms of 
the sneutrino mass come from the lepton-number conserving interactions, they 
should not contribute to the Majorana mass of a neutrino.  Only the 
lepton-number violating sneutrino mass, which is the mass-squared difference, 
should contribute to the Majorana neutrino mass.  This makes the neutrino 
mass proportional to the mass-squared difference after GIM cancellation. 
Note that combining both one-loop diagrams, a two-loop diagram is obtained 
which is similar to the diagram proposed in Ref.~\cite{bfpt} with a 
different lepton-number violating soft term.

In the case of three families, the induced neutrino mass terms involving 
$\nu_e$ are suppressed by the small value of the electron mass with respect 
to the $\mu$ or $\tau$ mass.  Therefore, unless $N^{\prime e}_e$ is much 
larger than $N^{\prime e}_\tau$ and $N^{\prime e}_\mu$, $\nu_e$ essentially 
decouples and acquires a very small mass; we get $m_{\nu_e} \sim 10^{-8}$~eV 
or less.  In this case, the mass matrix of the sneutrinos in the $\mu - \tau$ 
sector is of the form:
\begin{equation}
{\cal L} \owns - \frac{1}{2} \Phi^\dagger_{\tilde \nu} {\cal M}^2_{\tilde \nu} 
\Phi_{\tilde \nu},
\end{equation}
with $\Phi_{\tilde \nu} = (\tilde \nu_\mu, \tilde \nu_\tau, 
\tilde \nu_\mu^\dagger, \tilde \nu_\tau^\dagger )^T$ and
\begin{equation}
{\mathcal{M}}_{\tilde \nu}^2=\left(
\begin{array}{cccc}
M^2_{{\tiny L}_\mu} & 
0 & 
\delta m^2_{\tilde \nu_{\mu \mu}} & 
\delta m^2_{\tilde \nu_{\mu \tau}} 
 \\
0 &
M^2_{{\tiny L}_\tau} & 
\delta m^2_{\tilde \nu_{\mu \tau}} & 
\delta m^2_{\tilde \nu_{\tau \tau}} \\
\delta m^2_{\tilde \nu_{\mu \mu}} & 
\delta m^2_{\tilde \nu_{\mu \tau}} &
M^2_{{\tiny L}_\mu} & 
0 \\
\delta m^2_{\tilde \nu_{\mu \tau}} & 
\delta m^2_{\tilde \nu_{\tau \tau}} &
0 &
M^2_{{\tiny L}_\tau}
\end{array} \right)\,,\label{massmatrixsnu}
\end{equation}
where for simplicity we have assumed in Eq.~(\ref{lsoft}) 
a diagonal matrix $M^2_L=diag(M^2_{{\tiny L}_\mu}, 
M^2_{{\tiny L}_\tau})$ and with
\begin{equation}
\delta m^2_{{\tilde \nu}_{ij}} \sim \frac{1}{8 \pi^2} \frac{\mu^2 \xi^2}{v^2}
 m_{l_i} m_{l_j} 
\frac{N^{\prime e}_i N^{\prime e}_j}{(N^{\prime e})^2} \, .
\label{mnutilde2}
\end{equation}
{}From the mass matrix of Eq.~(\ref{massmatrixsnu}), the diagrams of Fig.~6 
induce then the following neutrino mass term
\begin{equation}
{\cal L} \owns - \frac{1}{2} \Psi^\dagger_{\nu} {\cal M}_{\nu} 
\Psi_{\nu},
\end{equation}
with $\Psi_\nu = (\nu_\mu, \nu_\mu^\dagger, \nu_\tau, \nu_\tau^\dagger)^T$ and
\begin{equation}
{\mathcal{M}}_{\nu} \sim A \, \left(
\begin{array}{cccc}
0 & 
\Delta_{\mu - \mu} &
0 & 
\Delta_{\mu - \tau} \\
\Delta_{\mu - \mu} &
0 & 
\Delta_{\mu - \tau} &
0 \\
0 &
\Delta_{\mu - \tau} & 
0 &
\Delta_{\tau - \tau} \\
\Delta_{\mu - \tau} & 
0 &
\Delta_{\tau - \tau} &
0
\end{array} \right)\,,\label{massmatrixnu}
\end{equation}
where
\begin{eqnarray}
\Delta_{i - i}&=& 
\MWp \Big( f( M^2_{{\tiny L}_i} +
\delta m^2_{\tilde \nu_{ii}}, \MWp^2 , 0)
- f( M^2_{{\tiny L}_i} - \delta m^2_{\tilde \nu_{ii}}, \MWp^2 , 0) \Big)
\nonumber\\ 
&\sim&
2 \delta m^2_{\tilde \nu_{ii}} 
\MWp  
\frac{
 M^2_{{\tiny L}_i} -\MWp^2- \MWp^2 \ln \Big( 
 M^2_{{\tiny L}_i}  / \MWp^2 \Big) }{\Big(
 M^2_{{\tiny L}_i} - \MWp^2 \Big)^2} ,
\\ 
\Delta_{\mu - \tau}&=& 
\frac{2 \delta m^2_{\tilde \nu_{\mu \tau}}}
{M^2_{{\tiny L}_\tau} - M^2_{{\tiny L}_\mu}} 
 \MWp 
\Big( f( M^2_{{\tiny L}_\tau} 
, \MWp^2 , 0)
- f( M^2_{{\tiny L}_\mu}, \MWp^2 , 0) \Big) \nonumber \\ &\sim& 
2 \delta m^2_{\tilde \nu_{\mu \tau}}
 \MWp (M^2_{{\tiny L}_\tau} - M^2_{{\tiny L}_\mu})^{-1} 
( M^2_{{\tiny L}_\tau} - \MWp^2)^{-1} 
( M^2_{{\tiny L}_\mu} -\MWp^2)^{-1}  \nonumber \\  
&\times& \left[ M^2_{{\tiny L}_\mu} \MWp^2 \ln \left( \displaystyle{
M^2_{{\tiny L}_\mu} \over \MWp^2 } \right)+ M^2_{{\tiny L}_\tau} 
 \MWp^2 \ln \left( \displaystyle{\MWp^2 \over M^2_{{\tiny L}_\tau} }
\right) + M^2_{{\tiny L}_\tau} M^2_{{\tiny L}_\mu} \ln \left( \displaystyle{
M^2_{{\tiny L}_\tau} \over M^2_{{\tiny L}_\mu} } \right) \right],
\end{eqnarray}
and
\begin{equation}
A=\frac{1}{64 \pi^2} \frac{e^2}{\sin^2 \theta_W}.
\end{equation}

This matrix can lead easily to a maximal mixing between 
the $\mu$ and $\tau$ neutrinos. This will be the case in particular
if $\Delta_{\mu - \mu} \sim \Delta_{\tau - \tau}$ 
which implies
\begin{equation}
\frac{m_\tau N^{\prime e}_\tau}
{M^2_{{\tiny L}_\tau}} \sim
\frac{m_\mu N^{\prime e}_\mu}
{M^2_{{\tiny L}_\mu}}.
\end{equation}
In addition it can be seen easily that in the limit where $M^2_{{\tiny L}_\mu}
 = M^2_{{\tiny L}_\tau}$, the determinant of the neutrino mass matrix 
vanishes, leading to a large hierarchy of masses (as required by atmospheric 
and solar neutrino experiments, taking into account the fact that the mass of 
$\nu_e$ is below $10^{-8}$~eV in the present scenario).  For example with the 
parameters of Eq.~(\ref{set1}) and taking in addition $N^{\prime e}_\mu =14 
N^{\prime e}_\tau$, $M_{{\tiny L}_\mu} \sim \MBp = 6$~TeV, and 
$M_{{\tiny L}_\tau}\sim 7.5$~TeV, we obtain one neutrino with a mass 
$\sim 10^{-3}$~eV and one with a mass $\sim 10^{-5}$~eV in addition to the 
electron neutrino with a mass below $10^{-8}$~eV.  In this case the mixing 
between the $\mu$ and $\tau$ flavors is large ($\sin 2 \alpha = 0.99$) while 
that of the electron flavor with the two other flavors is very much 
suppressed.  Note that the values of the lepton-number violating mass terms 
$\delta m^2_{\tilde \nu_{ij}}$ induced by Fig.~5 are several orders 
of magnitude below the phenomenological bounds $\delta m_e < 350$~MeV, 
$\delta m_\mu < 50$~GeV, and $\delta m_\tau < 450$~GeV obtained for 
$M_{\tiny SUSY} \sim 1$~TeV in Ref.~\cite{snu3}.


In summary, from the above qualitative estimate, we observe that realistic 
neutrino masses could be accomodated easily in the present scenario, in 
agreement with atmospheric and solar neutrino experiments.  A large mixing 
and a hierarchy of neutrino masses appear rather naturally.  A more 
quantitative estimate would require an explicit calculation of the two-loop 
integrals involved, but since there are still many free parameters, it will 
not add much to our understanding in any case.

\section{Conclusion}

We have studied a model of leptogenesis in a R-parity violating supersymmetric 
model.  The lightest neutralino $\tilde W'_3$ is assumed to be mostly the 
$SU(2)$ gaugino but its decay into $l^\pm h^\mp$ is suppressed because the 
required $\tilde l_L - h^+$ mixing is negligible.  On the other hand, $\tilde 
W'_3$ has a small component of $\tilde B$, the $U(1)$ gaugino, which decays 
readily because the required $\tilde l_R - h^+$ mixing is of order $10^{-3}$ 
from the presence of nonholomorphic R-parity violating soft terms in the 
Lagrangian.  The decay asymmetry of $\tilde W'_3$ is then evolved into a 
lepton asymmetry of the Universe by solving the Boltzmann equations in detail 
numerically.  We demonstrate how each term in the equations affects the 
eventual outcome of the proposed scenario.  The charged scalar mass matrix 
and the neutralino sector are discussed in detail.  A realistic scenario of 
radiative neutrino mass generation in two loops is presented, which 
originates from the same lepton-number violating nonholomorphic terms.  

\begin{center}{\large Acknowledgements}
\end{center}
This work was supported in part by the U.S.~Department of Energy under 
Grant No. DE-FG03-94ER40837.  One of us (U.S.) acknowledges the hospitality 
of the University of California at Riverside where this work was completed.

\newpage
\begin{appendix}
\section{Complete Charged Scalar Mass Matrix}

{}From Eqs.~(\ref{superpot}) to (\ref{lsoftNHRviol}), 
neglecting small terms of order $(\lambda_{ijk})^2$ 
and $\lambda_{ijk} {\lambda'}_{lmn}$, the complete 
charged scalar mass matrix is given by Eq.~(\ref{massmatrixdef})
with (see also Refs.~\cite{cf,mix} for the holomorphic part):
\begin{eqnarray}
{\mathcal{M}}^2_{{h_1^-}-{h_1^+}}&=&
\frac{g^2}{4}(v_2^2-v_{\nu}^2) 
-B \frac{v_2}{v_1}  
-\mu \mu_i \frac{v_{\nu_i}}{v_1} 
+\frac{1}{2}f^e_{ij} f^e_{kj} v_{\nu_i} v_{\nu_k} -N^{\prime {\tiny B}}_i
\frac{v_{\nu_i}}{v_1},
\label{h1h1}
 \\
{\mathcal{M}}^2_{{h_1^-}-{h_2^+}}&=&
 \frac{g^2}{4} v_1 v_2  -B,\\
{\mathcal{M}}^2_{{h_1^-}-{\tilde e_{i_{\tiny L}}}^*} &=&
\frac{g^2}{4} v_1 v_{\nu_i}
+\mu \mu_i -\frac{f^e_{kj} f^e_{ij}}{2} v_1 v_{\nu_k} 
+\frac{1}{2} f^e_{lj} (\lambda_{ikj} - \lambda_{kij}) v_{\nu_l} v_{\nu_k}
+N^{\prime {\tiny B}}_i ,\\
{\mathcal{M}}^2_{{h_1^-}-{\tilde e_i^c}}&=&
-\frac{1}{\sqrt{2}}f^e_{ji}  \mu_j v_2 
- \frac{1}{\sqrt{2}}(A_e f_e)_{ji} v_{\nu_j}  
+\frac{1}{\sqrt{2}} N^{\prime e}_i v_2,\\
{\mathcal{M}}^2_{{h_2^-}-{h_2^+}}&=&
\frac{g^2}{4} (v_1^2 + v_{\nu}^2) 
-B \frac{v_1}{v_2} 
\label{h2h2}
-B'_i \frac{v_{\nu_i}}{v_2}, \\
{\mathcal{M}}^2_{{h_2^-}-{\tilde e_{i_{\tiny L}}}^*} &=&
\frac{g^2}{4} v_2 v_{\nu_i} 
-B'_i , 
 \\
{\mathcal{M}}^2_{{h_2^-}-{\tilde e_i^c}}&=&
+\frac{1}{\sqrt{2}}f^e_{ji} (\mu v_{\nu_j} - \mu_j v_1)
+ \frac{1}{\sqrt{2}}\lambda_{kji} 
(\mu_k v_{\nu_j} - \mu_j v_{\nu_k} )\nonumber\\ 
&&+\frac{1}{\sqrt{2}} N^e_{ji} v_{\nu_j}+
\frac{1}{\sqrt{2}} N^{\prime e}_i v_1,\\
{\mathcal{M}}^2_{{\tilde e_{j_{\tiny L}}}-\tilde e_{i_{\tiny L}}^*} &=& 
(M^2_L)_{ij} -\frac{1}{8} (g^2-g'^2)(v_1^2-v_2^2+v_{\nu}^2) \delta_{ij} 
+\frac{g^2}{4} v_{\nu_i} v_{\nu_j} 
+\frac{1}{2}f^e_{jk} f^e_{ik} v_1^2 +\mu_j \mu_i 
\nonumber\\
&&+ \frac{1}{2} f^e_{jl}  v_1 v_{\nu_k} ( \lambda_{kil} -\lambda_{ikl}) 
+ \frac{1}{2}f^e_{il} v_1 v_{\nu_k} ( \lambda_{kjl} -\lambda_{jkl}), \\
{\mathcal{M}}^2_{\tilde e_{j_{\tiny L}}- \tilde e_i^c} &=& 
\frac{1}{\sqrt{2}}f^e_{ji} \mu v_2 + \frac{1}{\sqrt{2}} (A_e f_e)_{ji} {v_1}
+ \frac{1}{\sqrt{2}}(A^{\prime e}_{kji} -A^{\prime e}_{jki}) v_{\nu_k}\nonumber\\
&&+ \frac{1}{\sqrt{2}} \mu_k ( \lambda_{kji} - \lambda_{jki} ) v_2 
+\frac{1}{\sqrt{2}}N^e_{ji} v_2,
\\
{\mathcal{M}}^2_{{\tilde e_{j}^{c*}}-{\tilde e_{i}^c}}&=&
(M^2_{\tilde{e}^c} )_{ji} 
-\frac{g'^2}{4} (v_1^2-v_2^2+v_{\nu}^2) \delta_{ij}
+\frac{1}{2}f^e_{ki} f^e_{lj} v_{\nu_k} v_{\nu_l}
+\frac{1}{2}f^e_{ki} f^e_{kj} v_1^2  \nonumber\\
&&+\frac{1}{2} f^e_{kj} (\lambda_{lki} -\lambda_{kli}) v_1 v_{\nu_l} 
+\frac{1}{2}   f^e_{ki} (\lambda_{lkj} -\lambda_{klj}) v_1 v_{\nu_l}.
\label{ecec}
\end{eqnarray}
In Eqs.~(\ref{h1h1}) and (\ref{h2h2}),  
the tadpoles conditions for the $h_1^0$ and $h_2^0$ fields have been used:
\begin{eqnarray}
&& \Big( m^2_{H_1} + \mu^2 + \frac{1}{8}(g^2 + {g'}^2)
(v_1^2-v_2^2+v_{\nu}^2) \Big) v_1 +
(\mu \mu_i + N^{\prime {\tiny B}}_i) v_{\nu_i} \,\,=\,\,0,\\
&& \Big( m^2_{H_2} + \mu^2 + \mu_i^2 - \frac{1}{8}(g^2 + {g'}^2)
(v_1^2-v_2^2+v_{\nu}^2) \Big)v_2 +B v_1 + B'_i v_{\nu_i} \,\,=\,\,0.
\end{eqnarray}
In Eqs.~(\ref{h1h1}) to (\ref{ecec}), 
$v_{\nu_i}$ are given by the corresponding tadpole
conditions for the $\tilde{\nu_i}$ fields:
\begin{equation}
\Big( (M^2_L)_{ji} + \mu_i \mu_j +\frac{1}{8} (g^2 +{g'}^2) 
(v_1^2-v_2^2+v_{\nu}^2) \delta_{ij} \Big)v_{\nu_j} + B'_i v_2 + 
N^{\prime {\tiny B}}_i v_1 
+ \mu \mu_i v_1\,\,=\,\,0.
\end{equation}

Without nonholomorphic terms, it is difficult to obtain a large 
$\tilde{e}_R-h^+$ mixing without generating a large $\tilde{e}_L-h^+$ 
mixing as well (which would induce an undesirably large $\Wp \rightarrow 
h^+ e_L$ decay rate) or without requiring very fine tuning between the 
values of $v_{\nu_i}$ and $\mu_i$.  In the case of one family (putting all 
indices equal), this can be seen easily.  First, the 
${\mathcal{M}}^2_{{h_2^-}-\tilde e^c}$ matrix element is proportional to the 
neutrino mass and hence very small.  Second, the 
${\mathcal{M}}^2_{h_1^--\tilde e^c}$ matrix element, neglecting a small term 
proportional to the neutrino mass, is proportional to the 
${\mathcal{M}}^2_{\tilde e_L-\tilde e^c}$ matrix element.  Hence 
it can be shown easily that it is not possible to have a sufficiently large 
${\mathcal{M}}^2_{h_1^--\tilde e^c}$ matrix element (inducing $\xi$ of order 
$10^{-3}$) together with a sufficiently small $\tilde e_L-h^-$ mixing . The 
latter mixing gets a contribution $\sim \xi (\mu/\mu_l) 
[{\mathcal{M}}^2_{\tilde e_L-\tilde e^c}/ 
max({\mathcal{M}}^2_{\tilde e_L-\tilde e_L^\ast},
{\mathcal{M}}^2_{\tilde e^{c \ast}-\tilde e^c})]$.
Now, in the case of three families, due to the $A^{\prime e}$ terms in 
${\mathcal{M}}^2_{\tilde e_L-\tilde e^c}$, both matrix elements are not 
any more proportional and the $\tilde e^i_L-h^+$ mixings can be made as 
small as necessary independently of the value of $\xi$.  However, for 
values of $\tilde e_L$ and $\tilde e^c$ masses of the order 4 TeV or more 
(see section 4), a value of $\xi$ around $10^{-3}$ requires that the 
${\mathcal{M}}^2_{h_1^--\tilde{e}^c}$ matrix element is of order 
$10^{-3} \times($4 TeV$)^2 \simeq$ (125 GeV)$^2$ which 
implies very large values of 
$\mu_l$ and $v_{\nu_l}$.  Hence extreme fine tuning between the values of 
$\mu_i$ and $v_{\nu_i}$ is needed to obtain a small enough neutrino mass 
in Eq.~(\ref{massnu}).
\end{appendix}

\newpage
\bibliographystyle{unsrt}

\newpage

\begin{figure}[t]
\centerline{
\epsfxsize = \textwidth \epsffile{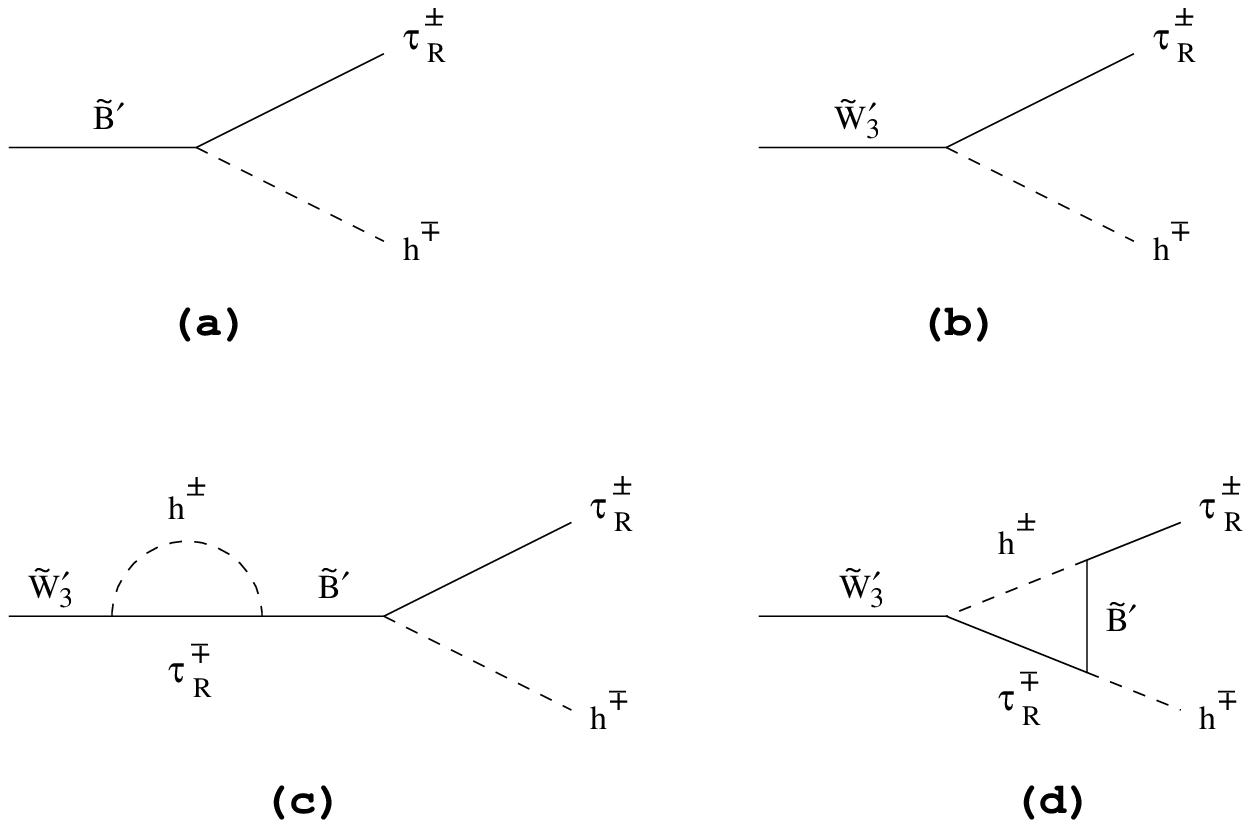}
}
\caption{  Tree-level diagrams for (a) $\tilde B'$
decay and (b) $\tilde W'_3$ decay (through their $\tilde B$ content),
and the one-loop (c) self-energy and (d) vertex diagrams for
$\tilde W'_3$ decay which have absorptive parts of opposite lepton number. }
\label{rpfg1}
\end{figure}
\begin{figure}[t]
\centerline{
\epsfxsize = \textwidth \epsffile{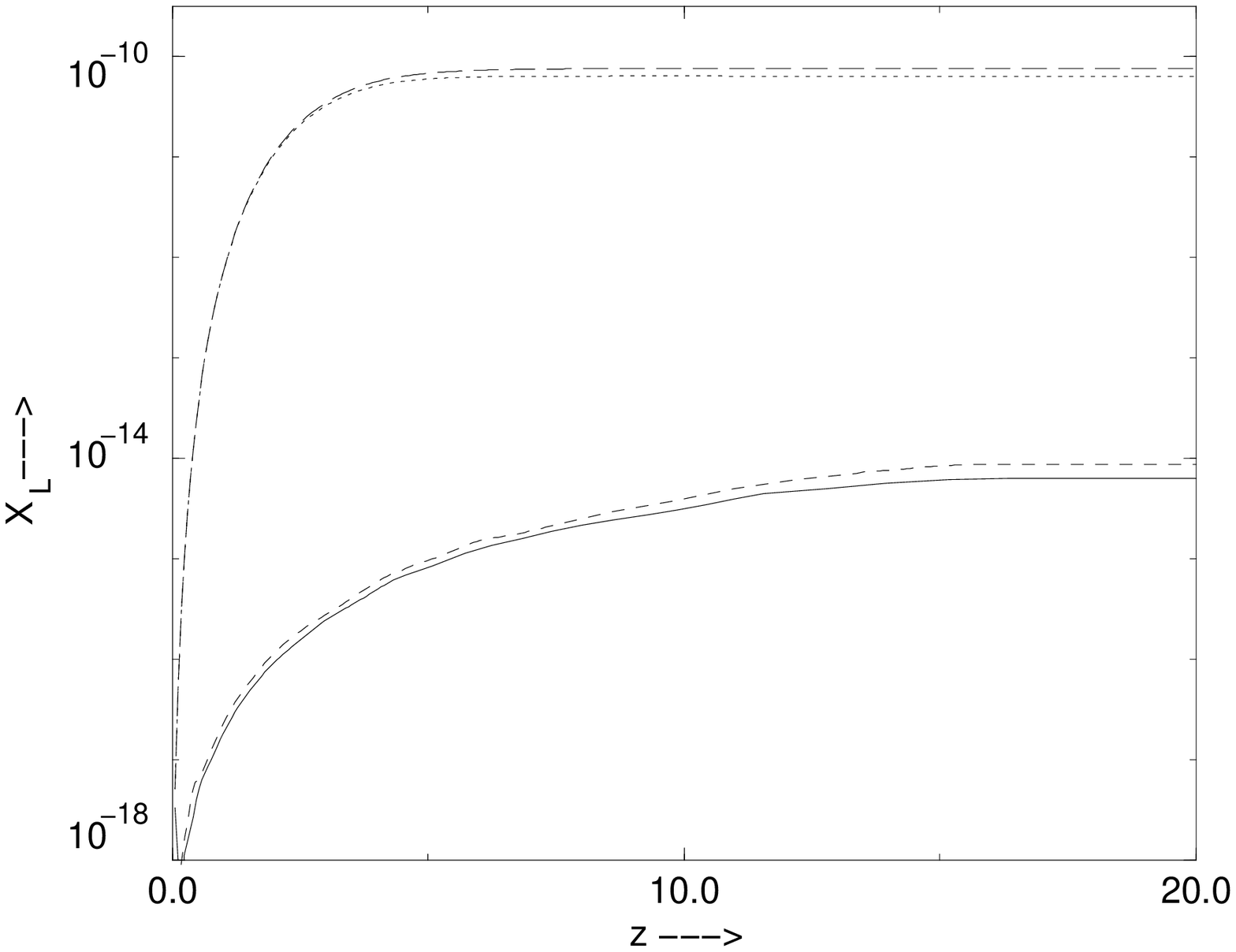}
}
\caption{ Leptonic asymmetry $X_L$ as a function of $z= \MWp/T$ as obtained
with the parameters given in the text, including all the contributions 
(solid); taking out the scattering term (short-dashed); not considering
in addition the inverse decay of the $\Bp$ damping term (dotted); and 
without the 
inverse decay of the $\Wp$ damping term (long-dashed). In the last case,
since all damping terms have been taken out, the asymptotic result is just
$X_L = \varepsilon n_\gamma /(2 s)$.  }
\label{rpfg2}
\end{figure}
\begin{figure}[t]
\centerline{
\epsfxsize = \textwidth \epsffile{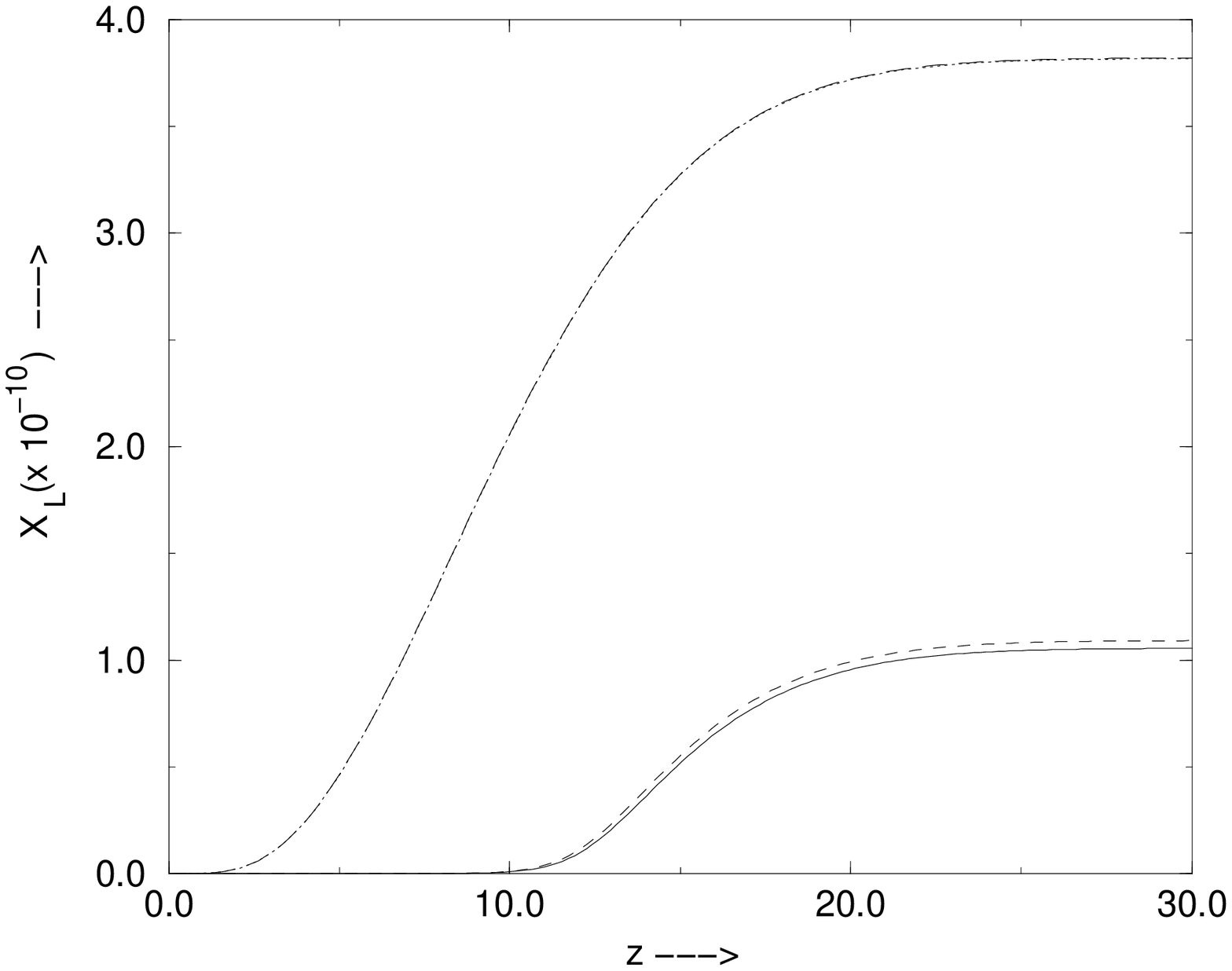}
}
\caption{Leptonic asymmetry $X_L$ as a function of $z= \MWp/T$ as obtained
with the parameters given by the set Eq.~(\ref{set1}), 
including all the contributions 
(solid); taking out the scattering term (short-dashed); not considering
in addition the inverse decay of the $\Bp$ damping 
term (dotted); and without the 
inverse decay of the $\Wp$ damping term (long-dashed). In the last case,
since all damping terms have been taken out, the asymptotic result is just
$X_L = \varepsilon n_\gamma /(2 s)$. }
\label{rpfg3}
\end{figure}
\begin{figure}[t]
\centerline{
\epsfxsize = \textwidth \epsffile{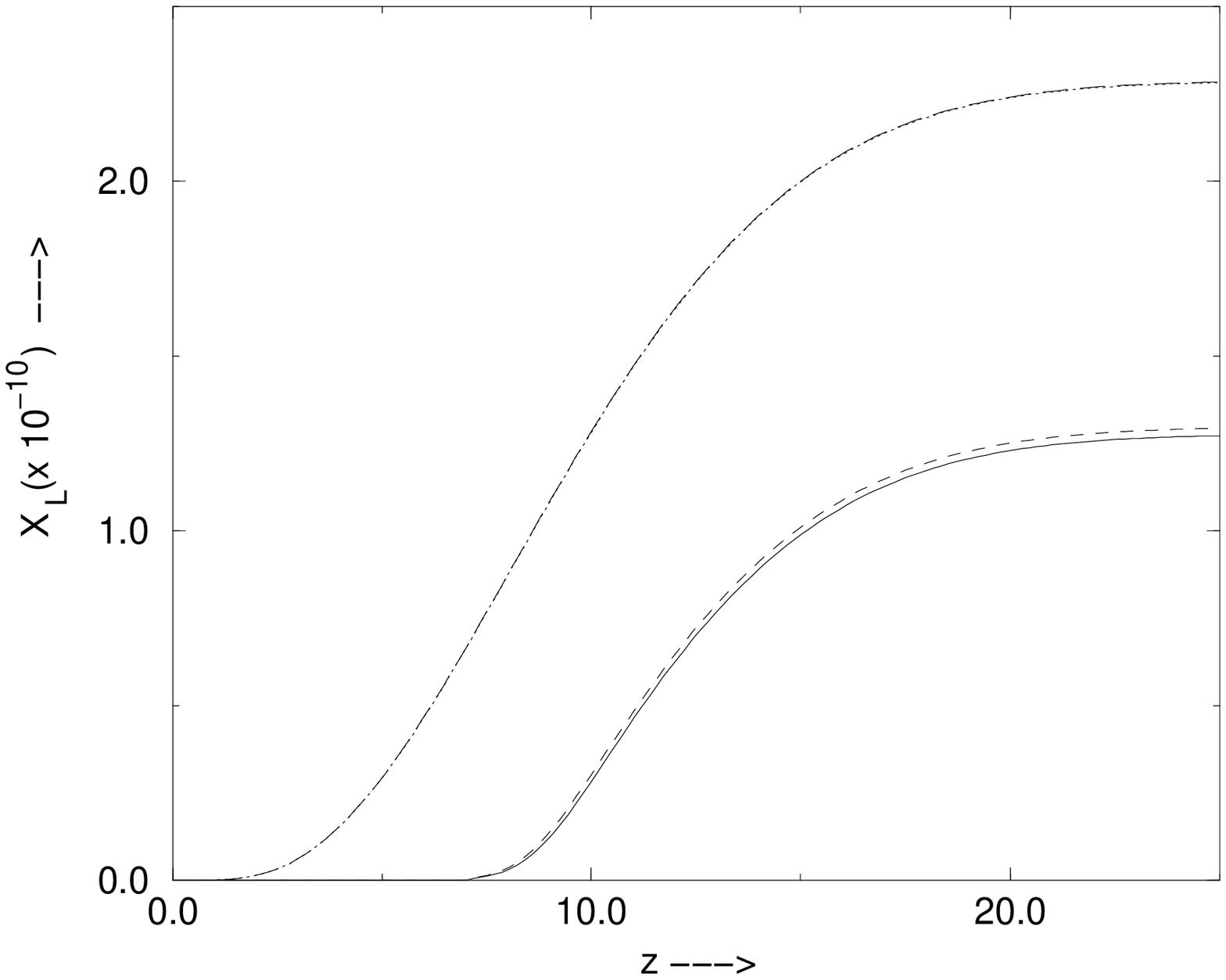}
}
\caption{Leptonic asymmetry $X_L$ as a function of $z= \MWp/T$ as obtained
with the parameters given by the set Eq.~(\ref{set2}), 
including all the contributions 
(solid); taking out the scattering term (short-dashed); not considering
in addition the inverse decay of the $\Bp$ 
damping term (dotted); and without the 
inverse decay of the $\Wp$ damping term (long-dashed). In the last case,
since all damping terms have been taken out, the asymptotic result is just
$X_L = \varepsilon n_\gamma /(2 s)$. }
\label{rpfg4}
\vskip 1in
\end{figure}
\begin{figure}[t]
\centerline{
\epsfxsize = 0.8\textwidth \epsffile{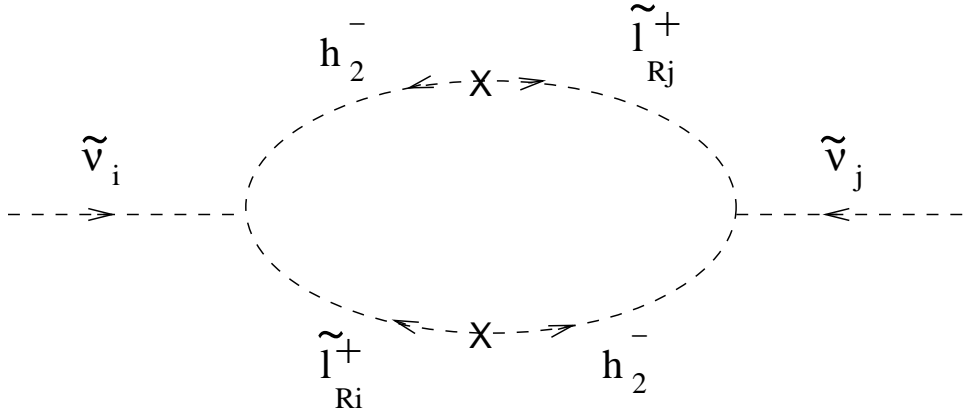}
}
\caption{One-loop diagram contributing to the sneutrino 
``Majorana'' mass.  }
\label{rpfg5}
\end{figure}
\begin{figure}[t]
\centerline{
\epsfxsize = 0.8\textwidth \epsffile{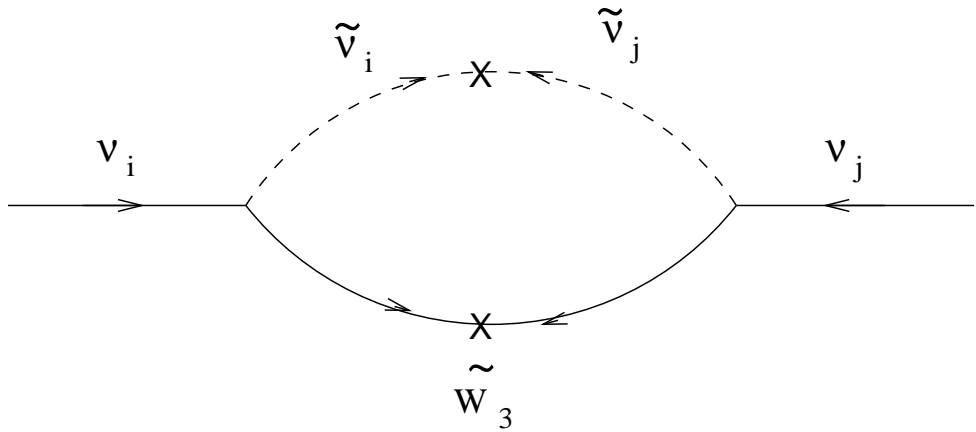}
}
\caption{One-loop diagram contributing to the neutrino mass, induced by
the sneutrino ``Majorana'' mass.  }
\label{rpfg6}
\end{figure}

\end{document}